\documentclass[preprint,12pt]{elsarticle}
\usepackage{amssymb}
\usepackage{amsthm}
\usepackage{hyperref}
\usepackage{color}
\journal{Optics Communication}

\begin{document}
\begin{frontmatter}
\title{Scattering of non-separable states of light}
\author[label1]{ P Chithrabhanu}\ead{chithrabhanu@prl.res.in}
\author[label1]{Salla Gangi Reddy}
\author[label1]{Ali Anwar}
\author[label1,label2]{A Aadhi}
\author[label1,label2]{Shashi Prabhakar}
\author[label1]{R P Singh}
\address[label1]{Physical Research Laboratory, Navrangpura, Ahmedabad. 380009, India.}
\address[label2]{IIT Gandhinagar, Chandkheda, Ahmedabad, India-382424 }
\begin{abstract}
We experimentally show that the non-separability of polarization and orbital angular momentum present in a light beam remains preserved under scattering through a random medium like rotating ground glass. We verify this by measuring the degree of polarization and observing the intensity distribution of the beam when projected to different polarization states, before as well as after the scattering. We extend our study to the non-maximally non-separable states also. 

\end{abstract}

\end{frontmatter}

\section{Introduction}
\label{Introduction}
A combined system is said to be entangled when its state cannot be expressed as a product of states corresponding to the individual sub systems \cite{Horodecki2009}. The entangled systems have interesting properties such as non-locality and contextuality which make them a great resource for various quantum protocols \cite{bouwmeester}. One generally uses the entanglement between two spatially separated particles in the same degree of freedom such as spin or polarization. However, one can also have hybrid entanglement in which two degrees of freedom of a single particle or two particles are entangled \cite{neves}. This arises due to the non-separability of two degrees of freedom. However, it is not an exclusive property of a quantum system. Similar kind of non-separability can be seen in classical optics, for example radially polarized light beams \cite{Zhan2009}. This quantum like classical entanglement has been receiving a lot of attention in recent years \cite{spreeuw1998,simon2010a, simon2, ghosh, Luis2009}. These non-separable states of light are shown to violate Bell like inequality \cite{borges,karimi2010}. Furthermore, they find applications in polarization metrology and ultra sensitive angular measurements \cite{toppel2,DAmbrosio2013}. 

 Recently, it has been shown that phase singular beams or optical vortices also violate Bell's inequality for continuous variables such as position and momentum \cite{gsa}. These optical vortices carry an orbital angular momentum (OAM) of $\pm l \hbar$ per photon,  $\pm l$  being the azimuthal index or order of the vortex \cite{allen, torres}.  This OAM can be used as an additional degree of freedom along with the polarization to form a hybrid entangled state that violates the Bell's inequality for discrete variables \cite{karimi2010}. 

 Scattering of structured light beams such as optical vortices has been studied for their coherence properties and applications \cite{ashok1, ashok2, ashok3, reddy1, lavery}. It has been shown that one can generate partially coherent ring shaped beams from the scattering of coherent optical vortices \cite{reddy2}. Here, we generate light beams with non-separable OAM and polarization and verify the preservation of non-separability under scattering through a rotating ground glass (RGG). These non-separable beams can be generated using q-plates \cite{Marrucci2006, Karimi2010a} or interferometers \cite{borges, Slussarenko2010a}. In our set up, we modify a polarizing Sagnac interferometer \cite{Slussarenko2010a} to generate the non-separable beams by replacing dove prism with a spiral phase plate (SPP). The generated beams scatter through a RGG and the scattered light is collected by a plano-convex lens to measure their polarization and intensity distributions at the focus. We measure the degree of polarization of the beam, as a measure of non-separability \cite{Gamel2012a, DeZela2014, Qian2011}, before and after scattering which should be $0$ for a maximally non-separable state and $1$ for a completely separable state. We also project the scattered as well as coherent light to different polarizations and record the corresponding intensity distributions which confirm the non-separability. Using the same experimental setup, we vary the degree of non-separability by controlling the intensities in the two arms of the interferometer. 

In section \ref{sec.2} we give a theoretical background to the OAM-polarization non-separable state and describe the methods we used to witness the non-separability. Experimental setup to generate the described states is given in section \ref{sec.3}. The results and discussion are given in section \ref{sec.4} and finally we conclude in section \ref{sec.5}. For simplicity, we use the Dirac notation to describe the states even though we are using classical light beams.
  \section{Theoretical Background}
\label{sec.2}
 A maximally entangled/non-separable state of polarization and OAM can be written as  
 \begin{equation}
  \vert \psi\rangle = \frac{1}{\sqrt{2}}\left(\vert H\rangle \vert +l\rangle + \vert V \rangle \vert -l\rangle\right)\label{1}
\end{equation}
 where $\vert H\rangle, \vert V \rangle $ and $\vert +l\rangle, \vert -l \rangle $ are basis vectors of 2D complex vector spaces corresponding to the polarization and the OAM subspace respectively. We work in the paraxial domain with linear optics, where polarization and OAM are independent. Thus \{$\vert H\rangle, \vert V \rangle $\} and \{$\vert +l\rangle, \vert -l \rangle $\} form two mutually independent complex vector spaces. The density matrix for the non-separable state $\vert\psi\rangle$ is given by $\rho_{ns} = \vert\psi\rangle\langle\psi\vert $. One can obtain the reduced density matrix corresponding to the polarization $\rho_p$ by taking a partial trace of this density matrix over OAM states,
\begin{equation}
\nonumber \rho_p = Tr_l\lbrace \rho_{ns}\rbrace = \sum_{i= l,-l} \langle i\vert\psi\rangle\langle\psi\vert i\rangle  \   = \frac{I_p}{2}.
\end{equation}
   Here, $I_P$ is a $2 \times 2$ identity matrix. For a given density matrix $\rho$ describing a state in $d$ dimensional Hilbert space, one can define linear entropy \cite{Peters}
   \begin{equation}
   S_L = \frac{d}{d-1}(1-Tr(\rho^2)).
   \end{equation}
$S_L$ characterizes the amount of mixedness for a given density matrix. It is known that for an entangled/non-separable state, the subsystems will be in a mixed state. Stronger the non-separability, larger the amount of mixedness present in the subsystems. Thus by measuring linear entropy $S_L$ of the subsystem, one can measure the degree of entanglement or the non-separability. For the maximally non-separable state given in Eq.~\ref{1}, one can find the linear entropy of polarization,
\begin{equation}
S_L = 2(1-Tr(\rho_p^2))= 1.\label{5}
\end{equation}
 This corresponds to a completely mixed polarization state in contrast to a completely polarized state with $S_L=0$.  We know, the state of polarization represented by a Poincare sphere can be completely described by
 \begin{equation}
 \rho_p = \frac{1}{2}\sum_{i=0}^{3} \sigma_i.s_i
 \end{equation} 
where $\sigma_i$'s and $s_i$'s are the Pauli matrices and normalized Stokes parameters respectively. The trace of square of this density matrix is given by 
\begin{equation}
Tr\lbrace\rho_p^2\rbrace = \frac{1}{2}\left(1+s_1^2+s_2^2+s_3^2\right) = \frac{1}{2}(1+DOP^2)\label{4}
\end{equation}
 where $DOP$ is the degree of polarization which is measured as the magnitude of the Stokes vector $\sqrt{s_1^2+s_2^2+s_3^2}$. Using Eq.~\ref{5} and Eq.~\ref{4} one can relate $DOP$ to the linear entropy,
 \begin{equation}
 S_L = 1-DOP^2.
\end{equation}   
  Thus for a maximally non-separable state of polarization and OAM, for which $S_L=1$, the degree of polarization should be zero. One can easily determine the DOP experimentally by measuring the Stokes parameters \cite{goldstein}.
   
   Another characteristic of the non-separable  state is the contexuality. For a separable state, measurement on one degree of freedom doesn't affect the measurement outcome of the other. However, in the case of a non-separable state, measurement outcome in one degree of freedom will depend on the context of measurement in the other. In our experiment the OAM state of the beam varies according to the projections to different polarization states due to their non-separability. Consider a general polarization state defined as
   \begin{equation}
   \vert \xi \rangle = Cos(\theta) \vert H \rangle + e^{i\phi} Sin(\theta) \vert V\rangle
   \end{equation}
   where $\theta$ and $\phi$ are the Euler angles corresponding to the state $\vert\xi\rangle$ on the Poincar\'e sphere. Projecting $\vert \psi\rangle$ given in Eq.~\ref{1} to $\vert\xi\rangle $, we obtain the OAM state as 
   \begin{equation}
 \vert\psi_o\rangle_{\theta,\phi} = \langle \xi\vert\psi\rangle 
 = Cos(\theta)\vert l \rangle + e^{-i\phi} Sin(\theta)\vert -l \rangle. 
   \end{equation}
This is a pure OAM superposition state. The transverse profile of the beam will correspond to the superposition of two equal and oppositely charged vortices with different relative amplitudes and phase. Therefore, the intensity profile of the beam varies according to the polarization projections defined by $\theta $ and $\phi $. For demonstration we take $(\theta, \phi) = (0,0), (90,0), (45,0), (-45,0),$ $(45,90)$ and $ (-45,90) $ which correspond to $\vert H\rangle, \vert V\rangle, \vert D\rangle = \vert H\rangle+\vert V\rangle , \vert A\rangle=\vert H\rangle-\vert V\rangle,\vert R\rangle=\vert H\rangle+i\vert V\rangle $ and $\vert L\rangle=\vert H\rangle-i\vert V\rangle $  polarization states. 
\begin{figure}[h]
\begin{center}

\includegraphics[scale=.48]{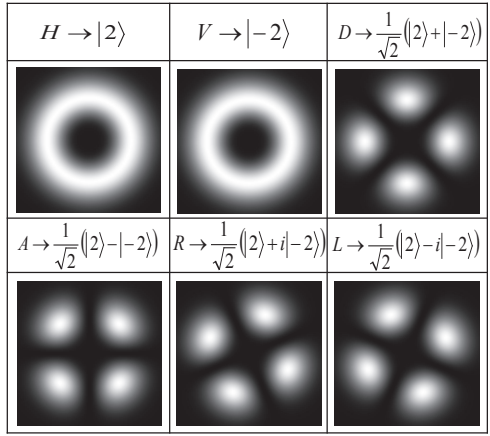} 
\caption{Theoretical images for the transverse intensity profile of a non-separable state described by Eq.~\ref{1} with $\vert l\vert =2$ for projections to different polarization states. H-Horizontal, V- vertical, D-diagonal, A-anit-diagonal, R-rightcircular, L-leftcircular}\label{fg.1}

\end{center}
\end{figure}

Figure~\ref{fg.1} shows the theoretical intensity distributions corresponding to different polarization projections for $\vert l\vert =2$. The projection on H (V) polarization gives a vortex of order $2 (-2)$. The projections of the state on diagonal (D), anti-diagonal (A), left circular (L) and right circular (R) gives superposition of two vortices that contain $2l$ (in our case $|l|=2$) number of lobes with different orientations. The number of lobes confirms the order or the azimuthal index of the vortex and the change in their orientation confirms the presence of non-separability in a light beam. 
 \section{Experiment}
\label{sec.3}
\begin{figure}[h]
\begin{center}
\includegraphics[scale=.47]{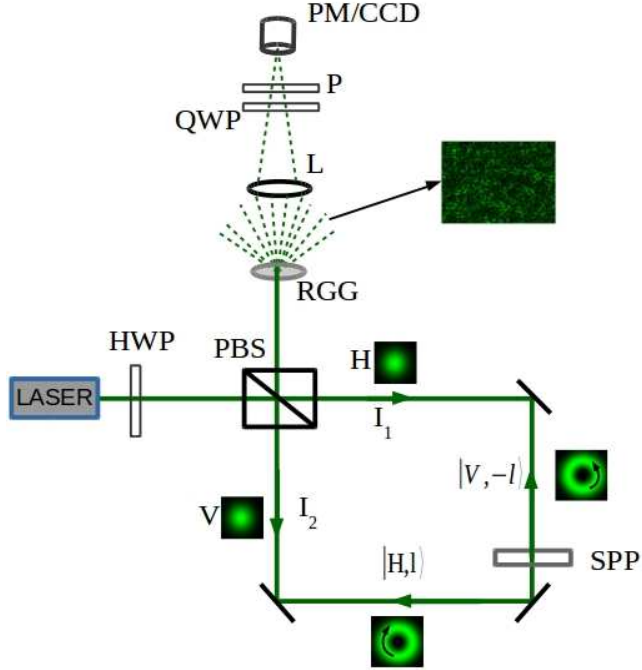} 
\caption{(Color online) Experimental setup for the generation and scattering of non-separable state of polarization and OAM. HWP- half wave plate, QWP- quarter wave plate, P- polarizer, $L$- lens with focal length 15 cm, CCD- charge coupled device (camera), PM-power meter, PBS- polarizing beam splitter}\label{fg.2}
\end{center}
\end{figure}

The experimental set up used to generate the non-separable state and to study its properties is shown in Fig.~\ref{fg.2}. We have used a diode pumped solid state green laser (Verdi 10) with vertical polarization for our study. The laser beam passes through a half wave plate, whose fast axis is oriented at $-22.5^o$ with the horizontal that changes beam polarization from vertical to diagonal. Then it passes through a polarizing Sagnac interferometer containing a spiral phase plate (SPP) to generate a light beam which is non-separable in polarization and OAM. 

 Two orthogonally  polarized (H and V) counter propagating Gaussian beams are converted into optical vortices of orders $l$ (for H) and $-l$ (for V) by the SPP designed for order $|l|=2$. These orthogonally polarized and oppositely charged vortices  superpose at the same PBS to form the described non-separable state. This non-separable state is generated only in the presence of SPP otherwise the superposition of two orthogonally polarized Gaussian beams results in a diagonally polarized Gaussian light beam. The doughnut shaped non-separable beam forms a random speckle distribution after scattering through the ground glass. A part of the scattered light collected with a lens of focal length 15 cm placed at a distance of 22 cm from the ground glass plate. The ground glass plate is rotating at $\approx$ 930 revolutions per minute to average out the speckles. The intensity distributions corresponding to the different polarization projections are recorded with an Evolution VF color cooled camera (pixel size $4.65 \mu$m) kept at the focus of the lens.

The Stokes parameters are measured using a quarter wave plate and a polarizer. We project the beam to horizontal (H), vertical (V), diagonal (D), anti-diagonal (D), right circular (R) and left circular (L) polarizations and measure the intensity. The intensity measurements for determining the Stokes parameters were performed with an optical power meter (Thorlab) of sensitivity 1 nW. One can find out the Stokes parameters as
\begin{eqnarray}
\nonumber s_1 &= \frac{I_H-I_V}{I};\\
s_2 &= \frac{I_D-I_A}{I};\\
\nonumber s_3 &= \frac{I_R-I_L}{I}
\end{eqnarray} 
where $I$ is the total intensity of the beam and $I_x$ is the intensity corresponding to $x$-polarization.

 \section{Results and Discussion}
\label{sec.4}
We have measured the Stokes parameters $(s_1, s_2, s_3)$ of coherent and scattered light beams for both separable (without SPP) and non-separable states (with SPP). We compare the degree of polarization of beams before and after scattering and the results are given in table~\ref{tb.1}. From the table, it is clear that the separable light beam is completely polarized (diagonal) while the non-separable state is completely unpolarized. The deviations in degree of polarization may be due to uncertainties in the orientation of the wave plates, small misalignment of the interferometer and the measurement uncertainty of the power meter. However, our experimental findings are very close to theoretical predictions given in section \ref{sec.2}.

\begin{table}
\begin{center}
 \begin{tabular}{|c|c|c|c|c|c|c|}
  \hline
   &\multicolumn{3}{|c|}{Before scattering}& \multicolumn{3}{|c|}{After scattering}\\ \cline{2-7}
   State & \multicolumn{2}{|p{1.3cm}|}{Stokes Vectors} &DOP& \multicolumn{2}{|p{1.3cm}|}{Stoke's Vectors} &DOP\\ \cline{1-7}
   Separable state& $s_1$ & 0.044 && $s_1$ & 0.056 &\\ \cline{2-3}\cline{5-6}(without SPP)& $s_2$ & 0.956 & 0.957 & $s_2$ & 0.922 &0.924\\  \cline{2-3}\cline{5-6}& $s_3$ & -0.02 & &$s_3$& -0.026&
\\ \hline
Non-separable& $s_1$ & -0.03 && $s_1$ & 0.01 &\\ \cline{2-3}\cline{5-6}state(with SPP)& $s_2$ & -0.01 & 0.001 & $s_2$ & -0.02 &0.001\\  \cline{2-3}\cline{5-6}&$s_3$&0.02& &$s_3$& -0.02&\\ \hline
\end{tabular} 
\caption{Stokes vectors and the degree of polarization corresponding to separable and non-separable states of light before and after scattering.}\label{tb.1}
\end{center}
\end{table}
We also generate non-maximally entangled states simply by controlling intensities in the two arms of the interferometer. This can be done easily by rotating the fast axis of the HWP. Then the state becomes
\begin{equation}
\vert \psi\rangle = \frac{1}{\sqrt{I_1+I_2}} \left(\sqrt{I_1}\vert H\rangle \vert +2\rangle + \sqrt{I_2} \vert V \rangle \vert -2\rangle\right)\label{2}
\end{equation}

\begin{figure}[h]
 \begin{center}
\includegraphics[scale=.6]{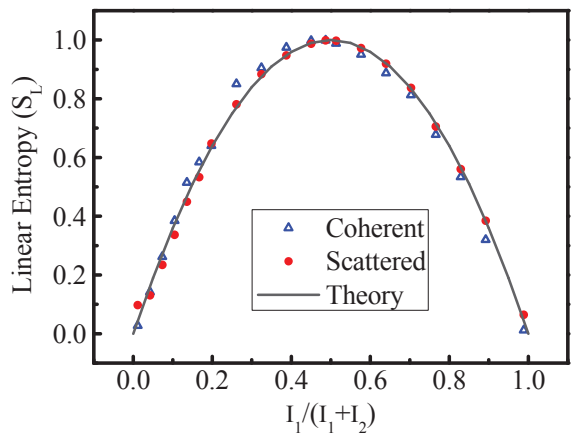} 
\caption{(Color online) Linear entropy \textit{vs.} normalized intensity $\frac{I_1}{I_1+I_2} $ plot for coherent and scattered non-separable states of light along with theoretical curve given by Eq.~\ref{3}.  }\label{fg.3}
\end{center}
\end{figure}

By varying $I_1 $ from 0 to $I$ and correspondingly $I_2$ from $I$ to 0, we have generated different states given in Eq.~\ref{2}. Note that the total intensity, $I_1+I_2 =I$ is always constant. For the state described in Eq.~\ref{2}, we can check the mixedness of the subsystem (here polarization) by calculating $S_L$ which also indicates the degree of non-separability. It reduces to a simple analytic expression,

\begin{equation}
S_L = \frac{4I_1I_2}{\left(I_1+I_2\right)^2}\label{3}.
\end{equation}

Line curve in Fig.~\ref{fg.3} shows the variation of linear entropy $S_L$ of polarization with the normalized intensity in one arm of the interferometer as given in Eq.~\ref{3}. The linear entropy becomes zero when $I_1 = 0$ or $I_2 = 0$, for which the state become  $\vert H\rangle\vert l\rangle$ and $\vert V\rangle\vert -l\rangle$ respectively. When the two intensities are same ($I_1 = I_2$), the state becomes completely non-separable for which $S_L = 1$.

We measure the Stokes parameters and calculate the degree of polarization and linear entropy experimentally corresponding to each value of $I_1 $ for coherent and scattered light beams. The results are shown in Fig.~\ref{fg.3}. One can clearly see that the $S_L$ \textit{vs.} normalized intensity curve for both the coherent and scattered light are in good agreement with the theoretical curve. The results of polarization measurements given in table~\ref{tb.1} and Fig.~\ref{fg.3} which confirm the preservation of non-separability in polarization and OAM under scattering by the RGG.
 \begin{figure}[t]
 \begin{center}
\includegraphics[scale=.43]{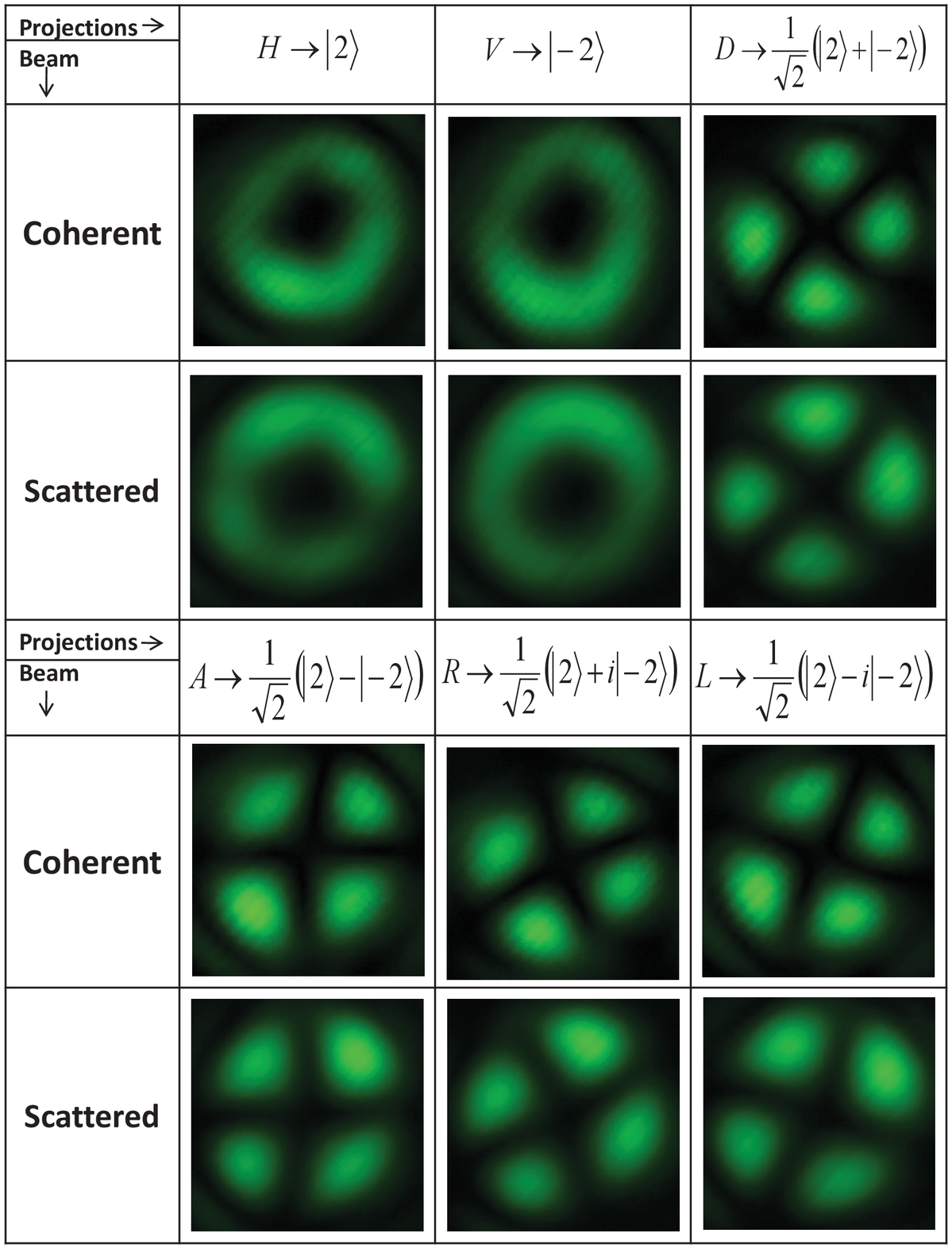} 
\caption{(Color online) Experimental images of coherent and scattered non-separable states of light with $l=2$ for different polarization projections. OAM states corresponding to each intensity distribution are also given. }\label{fg.5}
\end{center}
\end{figure}

Figure~\ref{fg.5} shows the intensity distributions for a coherent and a scattered light beam with non-separable state projected to the different polarizations. Our results show the similar behavior for both coherent and scattered light beams and are in good agreement with the theoretical images shown in Fig.~\ref{fg.1} that again confirm the preservation of non-separability. 
	
We also observe that the amount of scattered light collected by the lens is irrelevant regarding the non-separable properties. In fact, one can use multiple number of lenses  and collimate the scattered light again to form several copies of a partially coherent non-separable beam. This property can be used in public communication systems.
\section{Conclusions}
\label{sec.5}
In conclusion, we have produced a light beam with non-separable polarization and orbital angular momentum states using a simple interferometer and experimentally verified the preservation of the non-separability under scattering through a rotating ground glass.  The polarization measurements and the images of the beam projected to different polarizations show the presence of non-separability for coherent and scattered light. We have also demonstrated the generation of non-maximally non-separable states of light and studied their behavior under scattering by measuring the degree of polarization.





\end{document}